\documentclass[5p]{elsarticle}
\usepackage{amsmath,amssymb}
\usepackage{bbold}

\newcommand{\tr}[0]{\text{tr}}

\newcommand{\psibar}{\bar{\psi}}

\newcommand{\Bbar}{\bar{B}}

\newcommand{\path}[1]{[#1]}

 \allowdisplaybreaks[2]

\def\Dslash{\mathchoice
    {D\hskip-0.62em\raise0.2ex\hbox{$\displaystyle/$}\hskip0.2em}%
    {D\hskip-0.62em\raise0.2ex\hbox{$\textstyle/$}\hskip0.2em}%
    {D\hskip-0.5em\raise0.15ex\hbox{$\scriptstyle/$}\hskip0.2em}%
    {D\hskip-0.5em\raise0.15ex\hbox{$\scriptscriptstyle/$}\hskip0.2em}}

\newcommand{\DSGW}[0]{{\Dslash_{GW}}}

\newcommand{\DG}[0]{{\Dslash_{G}}}
\newcommand{\gL}[0]{\bar{\gamma}}
\newcommand{\gR}[0]{\hat{\gamma}}

\newcommand{\cbar}[0]{\overline{c}}
\begin{document}
\begin{frontmatter}
\title{The conserved axial current in the presence of multiple chiral symmetries}

\author[a]{Nigel Cundy}
\author[a]{Weonjong Lee} 
          
\address[a]{   Lattice Gauge Theory Research Center, FPRD, and CTP, Department of Physics \&
    Astronomy, Seoul National University, Seoul, 151-747, South Korea\\ 
        E-mail:{ndcundy@phya.snu.ac.kr}}
\let\today\relax
\date{March 20 2014}

\begin{abstract}
{In response to a recent work by Mandula, we investigate whether there are any
ambiguities in the expression for the pion mass resulting from multiple chiral
symmetries. If the conserved current for Ginsparg
Wilson chiral symmetries is calculated in the usual way, different expressions of the chiral
symmetry lead to different currents. This implies an ambiguity in the
definition of the pion 
and pion decay constant for all Ginsparg-Wilson expressions of the Dirac
operator, including the overlap operator on the lattice (although all these
currents would have the same continuum limit). We use a renormalisation group
mapping procedure to consider local chiral symmetry transformations for a
continuum Ginsparg-Wilson ``Dirac-operator." We find that this naturally leads
to an expression for the conserved current which is independent of which of the
Ginsparg-Wilson symmetries is chosen. We recover the standard expressions for
the massive Dirac operator, propagator, and chiral condensate. Our main conclusion is that, when the currents are properly constructed and consistently applied, no observable depends on which Mandula symmetry is used; at least in these continuum Ginsparg-Wilson theories. We will consider whether the same argument applies to lattice theories in a subsequent paper.

}
\end{abstract}
\end{frontmatter}
\section{Introduction}
It is currently believed that the chiral perturbation theory Lagrangian for lattice fermions which obey a Ginsparg-Wilson symmetry~\cite{Ginsparg:1982bj,Luscher:1998pqa,Hasenfratz:1998ri} is the same as that of the continuum theory, baring lattice artefacts in the low energy constants and some corrections due to the breaking of the full Lorentz group which emerge at a high order in the expansion~\cite{Bar:2005tu,Bar:2002nr}. Chiral perturbation theory requires that the left and right handed fermion fields transform according to $SU(2)_L \times SU(2)_R$ (we just consider the two flavour theory). However, these symmetry transformations do not apply in a Ginsparg-Wilson theory if we use the local chiral transformations which have previously been used~\cite{Kikukawa:1998py} to derive the conserved axial and vector currents with the Ginsparg-Wilson overlap lattice fermions~\cite{Neuberger:1998my,Neuberger:1997bg,Neuberger:1998fp,Narayanan:1993ss,Narayanan:1993sk}. 
It turns out that this is related to the additional curiosity uncovered recently by Mandula~\cite{Mandula:2007jt,Mandula:2009yd}: 
there is not just one, but an infinite number of Ginsparg-Wilson chiral symmetries, each (apparently) with a different conserved current. 

We show that these problems are related to the local symmetry transformation used to derive the current. Because the infinitesimal variation $\epsilon(x)$ does not commute with the Ginsparg-Wilson equivalent of $\gamma_5$, various distinct formulations of the local     transformation are consistent with the global transformation. The physical quantities, such as the conserved current, depend on the precise formulation of this transformation. So which local transformation do we use?

Practically, the only difference between the various possible currents is only a matter of having different O($a^2$) lattice artefacts. One can continue to use the currents defined in~\cite{Kikukawa:1998py} and ~\cite{Mandula:2007jt,Mandula:2009yd}, and after a continuum extrapolation and renormalisation one will reach the correct answer. The difficulty is in the details of the continuum extrapolation: what formula do we use, and how can we trust that it is reliable? The correct choice will almost certainly depend on the choice of current used in the simulation. We would usually hope to use a lattice chiral perturbation theory to control the extrapolation, but how would this account for the different O($a^2$) errors in the currents? It is plausible at first sight that the different choices of operator used to represent the current merely lead to different artefacts in the low energy constants, but if so this ought to be shown rather than assumed -- and on reflection it seems to us to be unlikely. In principle, if not in practice, the effective Lagrangian ought to be derivable from the QCD Lagrangian by a change of field variables and integration over unwanted degrees of freedom. This process will, after renormalisation, uniquely fix the low energy constants. There are only three pions in a two flavour Ginsparg-Wilson version of QCD. If the theory is to make sense as a spontaneously broken quantum field theory the definition of these pions in terms of fields should be unique -- the three mesonic particles whose mass drops to zero with the quark mass. With a unique definition of the pion fields, there ought to be one unique, correct, set of renormalised low energy constants in a given renormalisation scheme at non-zero lattice spacing. From this effective Lagrangian we obtain a single expression of the conserved current. So how can we reconcile this with multiple choices of the conserved current operator in lattice QCD? We would have to conclude that at most one of these is the correct operator corresponding to the sea fermions and the Symanzik chiral Lagrangian; and the others at best are the conserved currents obtained from a different effective action than that used for the sea quarks.

The first question to be asked, though, is if there are truly different possible representations of the conserved current in a Ginsparg-Wilson QCD, as implied by~\cite{Mandula:2007jt,Mandula:2009yd}. We propose here that this need not be the case -- if the currents are derived carefully. The multiple chiral symmetries are simply a by-product of the freedom to rotate the fermion and anti-fermion fields in the path integral formulation. We follow a procedure inspired by~\cite{Ginsparg:1982bj} which has been used in~\cite{Cundy:2009ab,Cundy:2010pu} and is related to (but differs from in several important respects) the procedure to construct the fixed point action~\cite{Hasenfratz:1994,Hasenfratz:1998ri}. When we rotate the fermion field, it alters the chiral symmetry operators, and this leads to the extended symmetry group uncovered by Mandula. However, the conserved current operators we find are not those used by Mandula; and as long as we are consistent (and use the same representation of the conserved current throughout the observable we are interested in -- to do otherwise in effect uses different fermion fields at different places in the operator, which leads to complications when contracting the fields) no physical observable depends on the choice of chiral symmetry -- as long as we use the `correct'\footnote{We use inverted commas around `correct' to make it clear to the reader what we mean by this term. All valid representations of the conserved current have the same continuum limit; so in this sense they are all correct. However, only one of them rightly represents the physics at non-zero lattice spacing for a given lattice Dirac operator; we describe this operator as `correct'.} representation of the conserved current. 

In this paper, we justify the claims in the above paragraph for a simple model in the continuum. Our reasons for doing so are: firstly, Mandula's symmetry group extends to all Ginsparg-Wilson theories, not just those on the lattice, so this is as good a place as any to test the consequences of the multiple chiral symmetries; secondly our methods are manifestly valid for the continuum theory while going to the lattice introduces more subtleties and complications which for an initial calculation we would rather avoid; thirdly that the answer in the continuum theory is well known and understood, and in particular there is a unique set of pions, and if we fail to find that we know that something has gone wrong; fourthly our methods and results might help guide us as we study Ginsparg-Wilson lattice theories. Of course, using this model has one major disadvantage that we do not intend to use it for any calculations; we still need to repeat the calculation for any Ginsparg-Wilson lattice theory used in numerical simulations, something which we address in a subsequent paper. Here, we will simply state that the idea that the multiple chiral symmetries of Mandula are related to each other by field rotations carries on trivially to the lattice: there is only one set of pions for the lattice theory as we would naively expect. However, identifying the `correct' representation of the conserved current on the lattice is more challenging than is often appreciated. However, here we simply discuss the methods which will be used in and motivate our companion paper.

This paper is organised as follows: in section \ref{sec:2}, we introduce the continuum model. In section \ref{sec:mappings} we discuss the field redefinitions used to link the different expressions of the Ginsparg-Wilson chiral symmetry and construct the conserved current. In section \ref{sec:standardcurrent} we outline the previous derivation of the various conserved currents associated with each chiral symmetry in a Ginsparg-Wilson theory, and why we believe it to be problematic. In section \ref{sec:5} we construct a current without the ambiguities of the standard approach. Section \ref{sec:po} discusses the implications for the expectation values of physical observables, and we conclude in section \ref{sec:conc}. 

Part of this work was previously presented in~\cite{Cundy:2011qe}.  
\section{Continuum Ginsparg-Wilson Dirac Operator}\label{sec:2}
In this work, we continually neglect the Yang-Mills part of the action as it does not affect any of our conclusions. We will consider the following Lagrangian with mass $m$ in the continuum in Euclidean Space Time,
\begin{multline}
\mathcal{L} = \psibar^{(1)} \DSGW \psi^{(1)} + \\ 
\psibar^{(1)} m(1-a\DSGW)(1+F(a^2\Dslash^2)) \psi^{(1)} + \cbar \DG c\label{eq:L1}
\end{multline}
$a$ is an (arbitrary) real dimensional parameter, $\psi^{(1)}$ and $\psibar^{(1)}$ are fermion fields -- which we distinguish from the fields $\psi$ and $\psibar$ of the standard continuum action $\psibar \Dslash \psi$ -- $\cbar$ and $c$ are fermionic ghost fields (with a mass of order $1/a$), and the operators are defined as\footnote{Throughout this work we use the notation that a real or complex constant, such as $a^{-1}$ in this equation, should be implicitly multiplied by the identity operator where appropriate.}
\begin{align}
\DSGW = & \frac{\Dslash (1 + F(a^2\Dslash^2))}{1 + a \Dslash (1 + F(a^2\Dslash^2))}\nonumber\\
\DG = & (a^{-1} +  \Dslash (1 + F(a^2\Dslash^2)))\frac{1}{1 + F(a^2\Dslash^2)}\label{eq:2}
\end{align}
$\Dslash = \gamma_\mu D_\mu$ is the massless continuum Dirac operator while $F(a^2\Dslash^2)$ is an arbitrary local real function of the Dirac operator whose eigenvalues are larger than $-1$. The ghost fields are introduced so that, after integration of the fermion and ghost fields, the partition function remains of the same form as the original continuum action.  Since we are only interested in those situations where $a$ is small (compared to the masses of the hadrons we want to study), the ghost will not affect any of the low energy dynamics, and we shall subsequently neglect it. $a$ is not a lattice spacing; just some arbitrary scale introduced into the theory.

 It is easy to confirm that $\DSGW$ is $\gamma_5$-Hermitian and satisfies the Ginsparg-Wilson relation~\cite{Ginsparg:1982bj}
\begin{gather}
 \gamma_5 \DSGW + \DSGW \gamma_5 (1-2a\DSGW) = 0.
\end{gather} 
If our only interest is the properties of Ginsparg-Wilson chiral symmetry, we may use $\DSGW$ as a simple test to investigate how we might expect a lattice Ginsparg-Wilson theory to behave. Our methods are more obviously applicable to this operator than on the lattice. As this Lagrangian is just to be used as a model to investigate the properties of the Ginsparg-Wilson chiral symmetry, we do not consider questions concerning its re-normalisability or perturbative expansion. 
The extension to the (more useful) lattice theory will be considered in a subsequent paper.

\section{Renormalisation group mappings}\label{sec:mappings}
The notation of this section follows~\cite{Cundy:2009ab,Cundy:2010pu}.

We start with a partition function for fermion fields $\psi_0$ and $\psibar_0$ (these are vectors containing all the flavours, so, for example, $\psibar_0 = (\psibar_{0u},\psibar_{0d})$) and gauge links $U$, constructed in terms of a Dirac operator $D_0$ (diagonal in flavour space) which is a function of $U$.
\begin{gather}
Z = \int d \psibar_0d\psi_0 dU e^{\psibar_0 D_0 \psi_0 }.
\end{gather}
Note that $\psi_0$ and $\psibar_0$ are not necessarily the fields $\psi$ and $\psibar$ of the original continuum action. This procedure is quite general, and we can use it on any action with any Dirac operator. 
We may choose to relate $D_0 = \Dslash$ and $\psi_0$ with the associated representation of the fermion fields, but we do not have to do so.
We construct a new partition function using the Ginsparg-Wilson mapping
procedure
\begin{multline}
Z = \int dU \int d\psibar_0 d\psi_0 e^{\psibar_0 D_0 \psi_0}\\
\int d\psibar_1 d \psi_1  e^{(\psibar_1 -
\psibar_0\Bbar^{-1})\alpha (\psi_1 - B^{-1}
\psi_0)},\label{eq:8a}
\end{multline}
where $B^{-1}$, $\Bbar^{-1}$ and $\alpha$ are some operators which define the blocking or
mapping. We require that $\alpha$ is invertible and $\det\alpha$ is independent of the fermion and gauge fields. If we block to a reduced
Hilbert space, then $B^{-1}$ and $\Bbar^{-1}$ are non-invertible operators (somewhat paradoxically given the notation) which
represent some sort of averaging procedure. For example, non-invertible
operators must be used to block from the continuum to the lattice. However, if we
map to the same Hilbert space (continuum to continuum, or lattice to a lattice of the same size), which is the only case we are interested in this work, then we may restrict ourselves to cases where
they are invertible, and the notation may be justified.

 In a discrete theory, these
mappings would therefore represent square matrices: we are not blocking or
averaging to, for example, reduce from the continuum to the lattice, but
constructing a different expression of the Dirac operator in the same
space-time. 

These mappings are integral transformations of the fields
\begin{gather}
(B^{-1}\psi)_a(x) \equiv \int d^4x' B^{-1}_{ab}(x,x') \psi_b(x'),
\end{gather}
for coordinates $x$ and spinor/colour indices $a$ and $b$. Throughout this work, we will only need to consider invertible mappings, where the inverse is defined as 
\begin{gather}
(B\psi)_b(x) \equiv \int d^4x' {B}_{ab}(x,x') \psi_b(x')
\end{gather}
The kernel ${B}_{ab}$ satisfies
\begin{gather}
\int d^4x' {B}_{ab}(x,x'){B}^{-1}_{bc}(x',x'') = \delta^{(4)}(x-x'') \delta_{ac}.
\end{gather}
 These mappings are functions of the gauge fields and contain a non-trivial
spinor structure. We then integrate over the fields $\psi_0$ to give a new
Lagrangian, $\mathcal{L}_1 = \psibar_1D_1\psi_1 + \tr\log[D + \Bbar^{-1}\alpha B]$. If it is not constant, the additional determinant,  $\det[D + \Bbar^{-1}\alpha B]$,
may be treated by introducing ghost fields, which
interact with the gauge fields but not (directly) with the fermions
and would describe a particle with a mass of the order of $a^{-1}$, where $a$
is the dimensional parameter introduced into the definition of $D_1$. This
additionally places the constraint on $B$ and $\Bbar$ that $[D + \Bbar^{-1}\alpha B]$ should be local. The mapped Dirac operator $D_1$ is
 \begin{align}
 D_1 =&\alpha - \alpha B^{-1} \frac{1}{\Bbar^{-1}\alpha B^{-1} + D_0}
\Bbar^{-1}\alpha \nonumber\\
=&  \alpha - \alpha \frac{1}{\alpha + \Bbar D_0 B}
\alpha.
 \end{align} 

 Suppose that $\gamma_{R0}$ and $\gamma_{L0}$ are some operators that generate a symmetry transformation of the fermion fields (for example, in the original continuum theory, for a vector chiral transformation we may choose $\gamma_{L0} = -1$ and $\gamma_{R0} = 1$, and for an axial chiral transformation we may choose $\gamma_{L0} = \gamma_{R0}=\gamma_5$), and that $\epsilon$ represents an infinitesimal parametrisation of the symmetry transformation (so in a U(1) transformation $\epsilon$ is a real number, and in an SU($n$) transformation $\epsilon$ is a Hermitian traceless $n\times n$ matrix). 
 Then if the original theory is invariant under an infinitesimal symmetry transformation
\begin{align}
\psi_0 \rightarrow & (1+i\epsilon \gamma_{R0}) \psi_0& \psibar \rightarrow& \psibar_0 (1+i\epsilon \gamma_{L0}),\label{eq:10}
\end{align}
then expanding the partition function in $\epsilon$ gives 
\begin{align}
Z &\rightarrow \int dU \int d\psibar_0 d\psi_0 e^{\psibar_0 D_0
\psi_0}\nonumber\\
&
\int  d\psibar_1 d \psi_1 e^{(\psibar_1 - \psibar_0(1+i\epsilon\gamma_{L0})\Bbar^{-1})\alpha (\psi_1 - B^{-1} (1+i\epsilon \gamma_{R0})\psi_0)}\nonumber\\
=& Z+i\int dUd \psibar_1 d \psi_1   \psibar_1\big(D_1B^{-1}
\gamma_{R0} \epsilon B+ \Bbar \gamma_{L0} \epsilon \Bbar^{-1}  D_1
-\nonumber\\
& D_1\alpha^{-1}\Bbar \gamma_{L0} \epsilon
\Bbar^{-1} D_1 - D_1 B^{-1} \gamma_{R0} \epsilon B \alpha^{-1}D_1\big)
\psi_1 e^{\psibar_1 D_1\psi_1}.\label{eq:gw1}
\end{align}
For a global transformation (where $\epsilon$ is constant), requiring that the partition function is invariant under chiral symmetry leads to a more generalised form of the Ginsparg-Wilson relation
\begin{multline}
D_1B^{-1} \gamma_{R0}  B+ \Bbar \gamma_{L0}  \Bbar^{-1}  D_1
-\\
D_1\alpha^{-1}\Bbar \gamma_{L0}  \Bbar^{-1} D_1 - D_1B^{-1}
\gamma_{R0} B \alpha^{-1}D_1 = 0.\label{eq:13}
\end{multline}
The procedure used to generate this relation is a generalisation of that used in~\cite{Ginsparg:1982bj}, who used the specific case where $\gamma_{L0} = \gamma_{R0} = \gamma_5$ and $[B,\gamma_5] = 0$ and $[\Bbar,\gamma_5]=0$. This is, then, simply a more general form of the familiar Ginsparg-Wilson relation.

In practice, there will be a family of mapping operators which generate the same Dirac operator $D_1$, which we parametrise as $B^{(\eta)}$, $\Bbar^{(\eta)}$ and $\alpha^{(\eta)}$. We will consider those mappings where $\alpha \rightarrow \infty \mathbb{1}$.

If $D_1$ and $D_0$ have the same eigenvectors with zero eigenvalue, the mapped theory will then obey a global Ginsparg-Wilson chiral symmetry defined by
\begin{align}
0= &\gL^{(\eta)} D_1 + D_1\gR^{(\eta)} & D_1 =& \Bbar^{(\eta)} D_0 B^{(\eta)}\nonumber\\
B^{(\eta)} \equiv& D_0^{-(\eta + 1)/2} D_1^{(\eta + 1)/2}&\Bbar^{(\eta)} \equiv & D_1^{(1-\eta)/2} D_0^{-(1-\eta)/2}.\nonumber\\
\gL^{(\eta)} \equiv &\Bbar^{(\eta)}\gamma_{L0}(\Bbar^{(\eta)})^{-1} &
\gR^{(\eta)} \equiv &   (B^{(\eta)})^{-1}\gamma_{R0} B^{(\eta)} ,\label{eq:egw}
\end{align}
Locality of $\gL^{(\eta)}$ and $\gR^{(\eta)}$ requires an odd integer
value of the parameter $\eta$. The choice of the chiral symmetry is equivalent
to the choice of $B$ and $\Bbar$ and therefore the choice of $\eta$. $\gamma_{L0}$ and $\gamma_{R0}$ represent the chiral symmetry operators associated with whichever fermion fields we are mapping from. So, if we apply an additional mapping to an already mapped theory (for example, if we want to apply a mapping to the fermion fields $\psi_1^{(1)}$ -- so we set $\psi_0$ in equation (\ref{eq:10}) to represent this field), we would substitute into equation (\ref{eq:egw}) the $\gL$ and $\gR$ of that mapped theory (for example by setting $\gamma_{L0} \equiv \gL^{(1)}$ and $\gamma_{R0}\equiv \gR^{(1)}$). So far, this argument is completely general, and can apply to many different fermion fields and Dirac operators. We later will apply it in various different specific situations.

First, we will map from a theory with continuum chiral symmetry to one with a Ginsparg-Wilson chiral symmetry (for example, an action built on the $\DSGW$ operator), so we must in this case choose $\gamma_{L0}$ and $\gamma_{R0}$ to be the generators of axial chiral symmetry in the continuum, i.e. $\gamma_{L0} =
\gamma_{R0} = \gamma_5$. In this case, if $D_1$ satisfies the standard Ginsparg-Wilson relation, we obtain~\cite{Cundy:2009ab}

If
mapping from a theory with continuum chiral symmetry, for those Dirac
operators which satisfy the standard Ginsparg-Wilson relation (i.e. whose
eigenvalues lie on a circle on the complex plane), then $\gamma_{L0} =
\gamma_{R0} = \gamma_5$ and~\cite{Cundy:2009ab}
\begin{align}
\gL^{(\eta)} = &(1-2D_1)^{\frac{1-\eta }{2}} \gamma_5&\gR^{(\eta)}= &\gamma_5 (1-2D_1)^{\frac{\eta + 1}{2}}\label{eq:15}
\end{align}

Note that it is impossible to construct the mappings from the chiral symmetry and Dirac operator.
For any
$\Bbar$ which generates the chiral symmetry transformation associated
with a given $\gL$ and $D_1$, one can construct another $\Bbar' = G\Bbar
F$ and $B' = F^{-1} BG^{-1}$ which generates the same chiral symmetry operator for any
invertible local operator $F$ which commutes with $\gamma_5$ and $D_0$ and any
invertible local operator $G$ which commutes with $\gamma_5$ and $D_1$.

For example, we may construct the Lagrangian (\ref{eq:L1}) by using the mapping 
\begin{align}
\Bbar^{(1)} =&1 & (B^{(1)})^{-1} =& (1+F) (1-a\DSGW),
\end{align}
and $D_0\equiv \Dslash$ and $D_1 \equiv \DSGW$,
which is valid as long as $1-\DSGW$ is invertible, which is true in this case: $(1-\DSGW)^{-1} = 1 + a \Dslash (1+F)$.
\section{Local Transformations: Standard Approach}\label{sec:standardcurrent}

For any $\gamma_5$-Hermitian Dirac operator, $D$, which commutes with its Hermitian
conjugate $[D,D^\dagger] = 0$, we can write
\begin{multline}
\left[\left(\frac{1}{-1} \frac{D+\zeta}{D^{\dagger}-\zeta}\right)^{\frac{1-\eta}{2}} \gamma_5 (D -
\zeta)\right.\\\left. + (D+\zeta) \gamma_5 \left(- \frac{D - \zeta}{D^\dagger
+\zeta}\right)^{\frac{1+\eta}{2}}\right]=0.\label{eq:66}
\end{multline}
We have written $\frac{1}{-1}$ rather than $-1$ to indicate how the root of this quantity should be taken (in comparison to the $(-1)^{\frac{1+\eta}{2}}$ in the second bracket) if $\eta$ is not odd integer.
$\zeta$ is some infinitesimal complex (or real) number, chosen only so that
$D^{\dagger} \pm \zeta$ is invertible (this is possible for Ginsparg Wilson
fermions, but not, for example, for Wilson lattice fermions where eigenvalues close to zero could take the values $|\zeta|e^{i\theta}$ for any $\zeta$ and $\theta$), and is included to ensure that there is a clear
and valid definition of the chiral symmetry operators as $\zeta \rightarrow
0$. In this limit, we can write that
\begin{gather}
\gL^{(\eta)} D + D \gR^{(\eta)} = 0
\end{gather}
with
\begin{align}
\gL^{(\eta)} =& \lim_{\zeta \rightarrow 0} \left(\frac{1}{-1}
\frac{D+\zeta}{D^{\dagger}-\zeta}\right)^{\frac{1-\eta}{2}}\gamma_5 \nonumber\\ 
\gR^{(\eta)} =&
 \lim_{\zeta \rightarrow 0} \gamma_5 \left(- \frac{D - \zeta}{D^\dagger
+\zeta}\right)^{\frac{1+\eta}{2}}.
\end{align}
These operators, which square to one, may be used to define chiral symmetry
transformations as long as they are local.
Obviously locality depends on the choice of $\eta$ and also on the Dirac operator:
in particular we assume that close to zero the allowed non-zero eigenvalues of
the Dirac operator are close to being purely imaginary, which will be true
for any Dirac operator with the correct limit as the momentum cut-off becomes
infinite (the correct continuum limit on the lattice; as above we expect
that a cut-off is incorporated in the definition of the Dirac operator). For
the overlap operator, these are local for odd integer $\eta$. For an anti-Hermitian
Dirac operator, such as the continuum operator, $\gL^{(\eta)} = \gR^{(\eta)} =
\gamma_5$. For any other normal Dirac operator, there will be multiple chiral
symmetries. 

Let us consider the case where $\eta = 1$ and $D$ is a Ginsparg-Wilson Dirac operator (satisfying the standard Ginsparg-Wilson equation $D + D^\dagger = 2 D^\dagger D$). In this case, we find 
\begin{align}
\gL^{(1)} = &\gamma_5 & \gR^{(1)} =& \gamma_5 (1-2aD),
\end{align}
which implies that the massless Lagrangian density $\psibar D \psi$ is invariant under the global transformation
\begin{align}
\psi \rightarrow & e^{i\epsilon \gR^{(1)}} \psi & \psibar \rightarrow  \psibar e^{i  \gL^{(1)} \epsilon},
\end{align}
and these are the standard L\"uscher Ginsparg-Wilson chiral symmetries~\cite{Luscher:1998pqa}.

The challenges arise when we try to convert this to a local symmetry. Most studies so far have used
\begin{align}
 \psibar \rightarrow & \psibar (1+{i  \gL^{(1)} \epsilon(x)})&\psi \rightarrow & (1+i\epsilon(x) \gR^{(1)}) \psi .\label{eq:localcurrent1}
\end{align}
However, there are several problems with this. Firstly, $[\epsilon,\gR^{(1)}] \neq 0$. This means that there are ambiguities when trying to define the local current. Do we choose $\psi\rightarrow (1+i\epsilon(x) \gR^{(1)})\psi$ or $\psi\rightarrow (1+i \gR^{(1)}\epsilon(x))\psi$ or something else\footnote{The original derivation of the current, \cite{Kikukawa:1998py}, appealed to the standard derivation of the Ginsparg-Wilson relation which uses $\alpha = 1$ and a $B$ and $\Bbar$ which commute with $\gamma_{R0}=\gamma_{L0}=\gamma_5$ \textit{and} $\epsilon(x)$ in equations (\ref{eq:gw1}) and (\ref{eq:13}). As shown above, this is not the only way to obtain the global Ginsparg-Wilson chiral symmetry via a mapping procedure, so the ambiguity is not resolved by this appeal to Ginsparg and Wilson.}? These will lead to different expressions of the conserved current. Secondly, we have the choice of $\eta$: each of these lead to a different conserved current~\cite{Mandula:2009yd}. Thirdly, the transformation of the left and right handed fields is not unitary. If we define a right handed fermion field as $\psi_R = \frac{1}{2}(1+\gR^{(1)}) \psi$ and a left handed field $\psi_L = \psi-\psi_R$, then the transformation of $\psi_R$ under a local vector (parametrised by $\epsilon_V$) and axial (parametrised by $\epsilon_A$) chiral transformation gives
\begin{align}
\psi_R \rightarrow& \frac{1}{2}(1+\gR^{(1)}) e^{i\epsilon_V(x) + i \epsilon_A(x) \gR^{(1)}}\psi\nonumber\\
=& e^{i\epsilon_V(x) + i \epsilon_A(x) \gR^{(1)}} \psi_R + \frac{1}{2} [\gR^{(1)},e^{i\epsilon_V(x) + i \epsilon_A(x) \gR^{(1)}}]\psi_R +\nonumber\\
& \frac{1}{2} [\gR^{(1)},e^{i\epsilon_V(x) + i \epsilon_A(x) \gR^{(1)}}]\psi_L.
\end{align}
This is particularly problematic when constructing a \\ Ginsparg-Wilson chiral perturbation theory, which requires that the transformations of $\psi_L$ and $\psi_R$ are within SU($N_f$): $\psi_L\rightarrow L \psi_L$, $\psi_R \rightarrow R \psi_R$. The conserved current is the generator of the pion fields; an ambiguous definition of the current implies that the pion field is ambiguous; and while these numerous pion fields are equivalent in the continuum limit, they still give us a conceptual headache away from the continuum, which may make us question if there is something deeply wrong with this whole understanding. 

\section{Local Transformations: Mapped approach}\label{sec:5}
Let us now return to the mapping procedure; and we are now studying the specific Ginsparg-Wilson Dirac operator defined in equation (\ref{eq:2}), so we set $D_1 \equiv \DSGW$. The original fermion field $\psi_0$ transformation is unambiguous
\begin{gather}
\psi_0 \rightarrow e^{i \epsilon_V + i \gamma_5 \epsilon_A} \psi_0,
\end{gather}
with $\epsilon_A,\epsilon_V \in su(N_f)$ for $N_f$ light fermion flavours\footnote{su($N_f$) represents the irreducible adjoint representation of the group, i.e. a Hermitian traceless matrix.}.
The mapped fermion field is $\psi_1^{(1)} = (B^{(1)})^{-1} \psi_0$, with $B^{(1)} = (1+F)(1-a\DSGW)$ so this transforms as
\begin{multline}
\psi_1^{(1)} \rightarrow \frac{1}{(1+F)(1-a\DSGW)} e^{i \epsilon_V + i \gamma_5 \epsilon_A}\\
 (1+F)(1-a\DSGW) \psi_1^{(1)}.\label{eq:t1}
\end{multline}
This definition is again unambiguous.
The right handed fermion field transforms in the same way,
\begin{align}
\psi_{1R}^{(1)} \rightarrow & \frac{1}{2} \left( 1 + \frac{1}{(1+F)(1-a\DSGW)} \gamma_5 (1+F)(1-a\DSGW)\right)\nonumber\\
&\phantom{space}\frac{1}{(1+F)(1-a\DSGW)}\ e^{i \epsilon_V + i \gamma_5 \epsilon_A} \nonumber\\
&\phantom{spacespac}(1+F)(1-a\DSGW) \psi_1^{(1)}\nonumber\\
=& \frac{1}{(1+F)(1-a\DSGW)} e^{i \epsilon_V + i  \epsilon_A} \nonumber\\&
\phantom{space}(1+F)(1-a\DSGW) \psi_{1R}^{(1)},
\end{align}
 which can be easily expressed in terms of the unitary operator $R = e^{i \epsilon_V + i \epsilon_A}$.
 Finally, the remaining chiral symmetries emerge by applying the additional mapping operation $B= (1-2\DSGW)^{-n}$ and $\Bbar = (1-2\DSGW)^{n}$ for integer $n$:
\begin{align}
\psi^{(1 + 4n)}_1 = & (1-2\DSGW)^n\psi^{(1)}_1 \nonumber\\
\psibar^{(1 + 4n)}_1 = & \psibar^{(1)}_1 (1-2\DSGW)^{-n}. \label{eq:mt}
\end{align}
Using (\ref{eq:egw}), this time setting $\gamma_{L0} = \gL^{(1)}$ and $\gamma_{L0} = \gR^{(1)}$ (since we are transforming from the $\psi^{(1)}$ basis to the $\psi^{(1+4n)}$ basis, we need to use the operators associated with the $\psi^{(1)}$ action), these lead to the chiral symmetry operators
\begin{align}
\gL^{(1 + 4n)} = &(1-2\DSGW)^n \gamma_5 (1-2\DSGW)^{-n} \nonumber\\
=& (1-2\DSGW)^{2n} \gamma_5 \nonumber\\
\gR^{(1+4n)} = &\gamma_5(1-2\DSGW)^{2n+1},  
\end{align}
in agreement with equation (\ref{eq:15}) if $\eta = 1+4n$. 

The axial and vector conserved currents may be defined as
\begin{gather}
J_{\mu}^{A,V} = \lim_{\epsilon \rightarrow 0}\frac{\partial}{\partial \partial_\mu \epsilon_{A,V}(x)} \delta_{\epsilon_A,\epsilon_V} \mathcal{L}
\end{gather} 
where $\delta_{\epsilon_A,\epsilon_V} \mathcal{L}$ indicates the change in the Lagrangian under a local chiral transformation. We find (for the $\eta = 1$ chiral symmetry)
\begin{multline}
\delta_{\epsilon_A,\epsilon_V} \mathcal{L}^{(1)} = \psibar^{(1)}_1 e^{-i\epsilon_V + i \gamma_5\epsilon_A}\DSGW \\ \frac{1}{(1+F) (1-a\DSGW)}e^{i\epsilon_V + i \gamma_5\epsilon_A}\\(1+F)(1-a\DSGW)\psi^{(1)}_1.
\end{multline}
Performing a Mandula transformation (equation (\ref{eq:mt})) gives
\begin{align}
\delta_{\epsilon_A,\epsilon_V} &\mathcal{L}^{(1 + 4n)} \nonumber\\
= &\psibar^{(1+4n)}_1 (1-2a\DSGW)^n \nonumber\\
&\phantom{spa}e^{-i\epsilon_V + i \gamma_5\epsilon_A}(1-2a\DSGW)^{-n}
\nonumber\\
&\phantom{spa}\DSGW (1-2a\DSGW)^n\frac{1}{(1+F) (1-a\DSGW)}
\nonumber\\
&\phantom{spa}e^{i\epsilon_V + i \gamma_5\epsilon_A}(1+F)(1-a\DSGW)\nonumber\\
&\phantom{spa}(1-2a\DSGW)^{-n}\psi^{(1+4n)}_1\nonumber\\
=& \psibar^{(1)}_1 e^{-i\epsilon_V + i \gamma_5\epsilon_A}\DSGW \nonumber\\
&\phantom{spa}\frac{1}{(1+F) (1-a\DSGW)}e^{i\epsilon_V + i \gamma_5\epsilon_A}\nonumber\\
&\phantom{spa}(1+F)(1-a\DSGW)\psi^{(1)}_1.
\end{align}
The current is therefore unchanged by the transformation. When we contract the $\psibar$ and $\psi$ fields we obtain an inverse Dirac operator which commutes with the factors of $(1-2a\DSGW)^{\pm n}$: these factors will not appear in any observables, as discussed below.

The pseudo-scalar and scalar operators are generated from the mass term in the Lagrangian, and may be calculated from
\begin{gather}
\{p,s\} = \frac{\partial}{\partial \epsilon_{\{A,V\}}(x)} \delta_{\epsilon_A,\epsilon_V} \mathcal{L}_M.
\end{gather}
The mass term for the Ginsparg-Wilson Lagrangian, $\mathcal{L}_M$, carries an additional factor of $(1+F)(1-a\DSGW)$ (easily derived by applying the mapping transformation to the massive Lagrangian), and may be defined as
\begin{gather}
\mathcal{L}_M=\psibar^{(1)}m (1+F)(1-a\DSGW) \psi^{(1)}_1.
\end{gather}
After the transformation, given that $[F,\DSGW]=0$, we obtain
\begin{align}
\delta_{\epsilon_A,\epsilon_V} \mathcal{L}_M =& \psibar^{(1)}_1 e^{-i\epsilon_V + i \gamma_5 \epsilon_A}m(1+F)(1-a\DSGW)  \nonumber\\
&\phantom{space}\frac{1}{(1+F)(1-a\DSGW)} e^{i\epsilon_V + i \gamma_5 \epsilon_A}\nonumber\\
&\phantom{spacespac}(1-a\DSGW)(1+F) \psi^{(1)}_1\nonumber\\
=& \psibar^{(1)}_1 e^{-i\epsilon_V + i \gamma_5 \epsilon_A}me^{i\epsilon_V + i \gamma_5 \epsilon_A}\nonumber\\
&\phantom{space}(1-a\DSGW)(1+F) \psi^{(1)}_1
\end{align}
and it is a straight-forward matter to show that change in $\mathcal{L}_M$ under these chiral transformations is invariant under the Mandula transformations. 

In this work, we have only considered the symmetries with $\eta = \ldots,1,5,9,\ldots$. Similar considerations apply when considering the symmetries for $\eta = \ldots,-1,3,7,\ldots$: these can be mapped to the $\eta = -1$ case by a field transformation, and a straight-forward calculation shows that the currents and (pseudo-)scalar operators derived from the $\eta = \pm 1$ symmetry transformations are equivalent.
\section{Physical observables}\label{sec:po}
The key element when deciding whether two theories are equivalent is to see whether the expectation values from those observables are the same. For simplicity, we just consider the axial current-current correlator, although the same arguments apply for all fermionic observables. We have
\begin{multline}
 J^{A,(n)}_\mu(z) =\\ \psibar^{(1+4n)} (1-2a\DSGW)^n \gamma_5 I(z) \gamma_\mu (1+F)\\(1-a\DSGW) (1-2a\DSGW)^{-n} \psi^{(1+4n)},
\end{multline}
with $I(z)_{xy} = \delta^{(4)}(z-x) \delta^{(4)}(z-y)$ the insertion operator at position $z$.
Thus the time ordered expectation value of the correlator gives,
\begin{multline}
\langle T[J_\mu^{A,(n)}(x) J_\nu^{A,(n)}(y)]\rangle = \theta(x^0-y^0) \langle J_\mu^{A,(n)}(x) J_\nu^{A,(n)}(y) \rangle \\+  \theta(y^0-x^0) \langle  J_\nu^{A,(n)}(y)J_\mu^{A,(n)}(x) \rangle
\end{multline}
Where $T$ represents the time ordering operator.\footnote{Time ordering adds a few additional subtleties which space forbids us from discussing in detail in this work. The original continuum fields ought to be time ordered, rather than the mapped Ginsparg-Wilson fields. It is best to expand the time-ordered correlation function in the continuum in terms of $\theta$-functions, and then perform all the various mappings to get the Ginsparg-Wilson theory in terms of non-time ordered correlation functions. Without time ordering, the propagator commutes with the Dirac operator. This will be discussed in more detail in the next work. }
At tree level,
\begin{align}
 \langle  J^{A,(n)}_\mu&(x)  J^{A,(n)}_\nu(y)\rangle  \nonumber\\
 =& \big\langle \psibar^{(1+4n)} (1-2a\DSGW)^n \gamma_5 I(x) \gamma_\mu (1+F)\nonumber\\
 &\phantom{spacespace}(1-a\DSGW) (1-2a\DSGW)^{-n} \psi^{(1+4n)}\nonumber\\
 &\phantom{spa}\psibar^{(1+4n)} (1-2a\DSGW)^n \gamma_5 I(y) \gamma_\nu (1+F)\nonumber\\
 &\phantom{spacespace}(1-a\DSGW) (1-2a\DSGW)^{-n} \psi^{(1+4n)} \big\rangle\nonumber\\
 =& \tr \bigg(\frac{(1+F)(1-a\DSGW)}{\DSGW} \gamma_5 \gamma_\mu I(x)\nonumber\\
 &\phantom{spa}\frac{(1+F)(1-a\DSGW)}{\DSGW} \gamma_5 \gamma_\nu I(y)\bigg) -\nonumber\\
 &\tr\bigg(\frac{(1+F)(1-a\DSGW)}{\DSGW} \gamma_5 \gamma_\mu I(x)\bigg)\nonumber\\
 &\phantom{spa}\tr\bigg(\frac{(1+F)(1-a\DSGW)}{\DSGW} \gamma_5 \gamma_\nu I(y)\bigg).
\end{align}
Clearly, this expression is independent of $n$. Note that these expressions are exactly what we obtain in standard QCD. There is nothing surprising about this: all we have done is rotate the fermion field, and have not altered the physical content of the theory in any way. When we include loops, the perturbative calculation (before renormalisation) will become complex, since we have to account for the ghost fields, $c$ and $\bar{c}$, and the more complicated form of the Dirac operator and the associated fermion/gauge vertex, but the final result will agree with that obtained from the standard continuum Dirac operator.

What if we considered $\langle \langle  J^{A,(n)}_\mu(x)  J^{A,(n')}_\mu(y)\rangle \rangle$, the correlator between two currents obtained from different chiral symmetries? What this analysis tells us is that we cannot naively contract $\psi^{(n)}$ with $\psi^{(n')}$; instead, we should use the relation between the fermion fields, $\psi^{(n')} = (1-2a\DSGW)^{n-n'} \psi^{(n)}$, which will allow us to contract like with like. The observables will remain independent of the choices of $n$ we select.

Obviously, if the observables are all in agreement then the same effective action can be used for each of the chiral symmetries.
\section{Conclusions} \label{sec:conc}
There are two conclusions to draw from this. Firstly, the multiple chiral symmetries are an artefact of our freedom to redefine the fermion field using a mapping procedure. This follows on the lattice just as much as in the continuum example considered in this letter. Secondly, if (and only if) the conserved current and fermion fields are defined correctly the physical quantities, the conserved current and pseudofermion operator, are independent of the choice of symmetry. There is only one set of pions in a continuum Ginsparg Wilson theory.

The principle purpose of this work is to emphasise the point that to avoid ambiguities in the observables in the Ginsparg-Wilson it is important to use the correct derivation of the conserved current, as derived from the RG mapping (or perhaps averaging) procedures. If the current is derived and constructed correctly, there are no ambiguities relating to Mandula's `multiple chiral symmetries'.  If some other observable is used to represent the conserved current, then it will not give a consistent result at non-zero lattice spacing; for example the observable will not be the correct observable to generate the pion field, but will give the pion field plus something else (or the pion field multiplied by some function of the pion's momentum). Another way of evaluating this situation might be to say that we are using the correct conserved current for a different lattice Dirac operator (if such a Dirac operator exists). This would lead to a mixed action theory with an unknown operator in the valance sector. While the difference is just a matter of lattice artefacts, it is still important because these effective field theories are used to derive the extrapolation formulae to control the continuum limit. If you use the wrong expression for the conserved current, then the standard  chiral perturbation theory for Ginsparg-Wilson fermions does not necessarily apply. We investigate whether the correct conserved current is presently being used in lattice simulations in the companion paper.

\section*{Acknowledgments}
Numerical calculations used servers at Seoul National University
funded by the BK21 program of the NRF (MEST), Republic of
Korea. NC was supported in part through the BK21 program of the NRF (MEST), Republic of Korea. This research was supported by Basic Science Research Program through the National Research Foundation of Korea(NRF) funded by the Ministry of Education(2013057640). W. Lee is supported by the Creative Research Initiatives
Program (2013-003454) of the NRF grant, and acknowledges the support
from the KISTI supercomputing center through the strategic support
program for the supercomputing application research
(KSC-2012-G2-01).

\bibliographystyle{elsarticle-num}
\bibliography{weyl}

\begin{thebibliography}{10}
\expandafter\ifx\csname url\endcsname\relax
  \def\url#1{\texttt{#1}}\fi
\expandafter\ifx\csname urlprefix\endcsname\relax\def\urlprefix{URL }\fi
\expandafter\ifx\csname href\endcsname\relax
  \def\href#1#2{#2} \def\path#1{#1}\fi

\bibitem{Ginsparg:1982bj}
P.~H. Ginsparg, K.~G. Wilson, A remnant of chiral symetry on the lattice, Phys.
  Rev. D25 (1982) 2649.

\bibitem{Luscher:1998pqa}
M.~L{\"u}scher, {Exact chiral symmetry on the lattice and the Ginsparg-{W}ilson
  relation}, Phys. Lett. B428 (1998) 342--345.
\newblock \href {http://arxiv.org/abs/hep-lat/9802011}
  {\path{arXiv:hep-lat/9802011}}.

\bibitem{Hasenfratz:1998ri}
P.~Hasenfratz, V.~Laliena, F.~Niedermayer, {The Index theorem in QCD with a
  finite cutoff}, Phys. Lett. B427 (1998) 125--131.
\newblock \href {http://arxiv.org/abs/hep-lat/9801021}
  {\path{arXiv:hep-lat/9801021}}, \href
  {http://dx.doi.org/10.1016/S0370-2693(98)00315-3}
  {\path{doi:10.1016/S0370-2693(98)00315-3}}.

\bibitem{Bar:2005tu}
O.~Bar, C.~Bernard, G.~Rupak, N.~Shoresh, {Chiral perturbation theory for
  staggered sea quarks and Ginsparg-Wilson valence quarks}, Phys.Rev. D72
  (2005) 054502.
\newblock \href {http://arxiv.org/abs/hep-lat/0503009}
  {\path{arXiv:hep-lat/0503009}}, \href
  {http://dx.doi.org/10.1103/PhysRevD.72.054502}
  {\path{doi:10.1103/PhysRevD.72.054502}}.

\bibitem{Bar:2002nr}
O.~Bar, G.~Rupak, N.~Shoresh, {Simulations with different lattice Dirac
  operators for valence and sea quarks}, Phys.Rev. D67 (2003) 114505.
\newblock \href {http://arxiv.org/abs/hep-lat/0210050}
  {\path{arXiv:hep-lat/0210050}}, \href
  {http://dx.doi.org/10.1103/PhysRevD.67.114505}
  {\path{doi:10.1103/PhysRevD.67.114505}}.

\bibitem{Kikukawa:1998py}
Y.~Kikukawa, A.~Yamada, {Axial vector current of exact chiral symmetry on the
  lattice}, Nucl. Phys. B547 (1999) 413--423.
\newblock \href {http://arxiv.org/abs/hep-lat/9808026}
  {\path{arXiv:hep-lat/9808026}}, \href
  {http://dx.doi.org/10.1016/S0550-3213(99)00059-0}
  {\path{doi:10.1016/S0550-3213(99)00059-0}}.

\bibitem{Neuberger:1998my}
H.~Neuberger, {A practical implementation of the overlap-Dirac operator}, Phys.
  Rev. Lett. 81 (1998) 4060--4062.
\newblock \href {http://arxiv.org/abs/hep-lat/9806025}
  {\path{arXiv:hep-lat/9806025}}.

\bibitem{Neuberger:1997bg}
H.~Neuberger, {Vector like gauge theories with almost massless fermions on the
  lattice}, Phys. Rev. D57 (1998) 5417--5433.
\newblock \href {http://arxiv.org/abs/hep-lat/9710089}
  {\path{arXiv:hep-lat/9710089}}, \href
  {http://dx.doi.org/10.1103/PhysRevD.57.5417}
  {\path{doi:10.1103/PhysRevD.57.5417}}.

\bibitem{Neuberger:1998fp}
H.~Neuberger, Exactly massless quarks on the lattice, Phys. Lett. B417 (1998)
  141--144.
\newblock \href {http://arxiv.org/abs/hep-lat/9707022}
  {\path{arXiv:hep-lat/9707022}}.

\bibitem{Narayanan:1993ss}
R.~Narayanan, H.~Neuberger, Chiral fermions on the lattice, Phys. Rev. Lett. 71
  (1993) 3251--3254.
\newblock \href {http://arxiv.org/abs/hep-lat/9308011}
  {\path{arXiv:hep-lat/9308011}}.

\bibitem{Narayanan:1993sk}
R.~Narayanan, H.~Neuberger, Chiral determinant as an overlap of two vacua,
  Nucl. Phys. B412 (1994) 574--606.
\newblock \href {http://arxiv.org/abs/hep-lat/9307006}
  {\path{arXiv:hep-lat/9307006}}.

\bibitem{Mandula:2007jt}
J.~E. Mandula, {Note on the Lattice Fermion Chiral Symmetry Group. }\href
  {http://arxiv.org/abs/0712.0651} {\path{arXiv:0712.0651}}.

\bibitem{Mandula:2009yd}
J.~E. Mandula, {Symmetries of Ginsparg-Wilson Chiral Fermions. }\href
  {http://arxiv.org/abs/0901.0572} {\path{arXiv:0901.0572}}.

\bibitem{Cundy:2009ab}
N.~Cundy, {A renormalisation group derivation of the overlap formulation},
  Nucl. Phys. B824 (2010) 42--84.
\newblock \href {http://arxiv.org/abs/0903.5521} {\path{arXiv:0903.5521}},
  \href {http://dx.doi.org/10.1016/j.nuclphysb.2009.08.016}
  {\path{doi:10.1016/j.nuclphysb.2009.08.016}}.

\bibitem{Cundy:2010pu}
N.~Cundy, {A Ginsparg-Wilson approach to lattice CP symmetry, Weyl and
  Majoranna fermions, and the Higgs mechanism}(Revised version in preparation).
\newblock \href {http://arxiv.org/abs/1003.3991} {\path{arXiv:1003.3991}}.

\bibitem{Hasenfratz:1994}
P.~Hasenfratz, F.~Niedermayer, Nucl. Phys. B414 (1994) 785.
\newblock \href {http://arxiv.org/abs/hep-lat/9308004}
  {\path{arXiv:hep-lat/9308004}}.

\bibitem{Cundy:2011qe}
N.~Cundy, W.~Lee, {Gell-Mann-Oakes-Renner relation for multiple chiral
  symmetries}, PoS (LAT2010) (2011) 249\href {http://arxiv.org/abs/1111.2638}
  {\path{arXiv:1111.2638}}.

\end{thebibliography}

\end{document}